\documentclass
[aps,prl,twocolumn,floatfix,english,showpacs,10pt,superscriptaddress]{revtex4-2}%
\usepackage{graphicx}
\usepackage{amsmath}
\usepackage{physics}
\usepackage{amssymb}
\usepackage{colordvi}
\usepackage{verbatim}
\usepackage{xcolor}
\usepackage{mathrsfs}
\usepackage{epsfig}
\usepackage{lipsum}
\usepackage{amsfonts}
\usepackage[unicode=true, breaklinks=false, pdfborder={0 0 1}, backref=false,
colorlinks=true, linkcolor=blue, urlcolor=blue, citecolor=blue]{hyperref}%
\providecommand{\U}[1]{\protect\rule{.1in}{.1in}}
\setcitestyle{numbers,square}
\begin{document}

\title{Chirality Enables Thermal Magnon Transistors}

\author{Tao Yu}
\email{taoyuphy@hust.edu.cn}
\affiliation{School of Physics, Huazhong University of Science and Technology, Wuhan 430074, China}
\author{Chengyuan Cai}
\affiliation{School of Physics, Huazhong University of Science and Technology, Wuhan 430074, China}
\author{Gerrit E. W. Bauer}
\email{G.E.W.Bauer@imr.tohoku.ac.jp}
\affiliation{WPI-AIMR and Institute for Materials Research and CSRN, Tohoku University, Sendai 980-8577, Japan}
 \affiliation{Kavli Institute for Theoretical Sciences, University of the Chinese Academy of Sciences, Beijing 100190, China}
 
\date{\today }

\begin{abstract}
We report a theory of thermal spin pumping into proximity magnets under a transverse-bias-driven heat flow of magnons in magnetic films when the dipolar coupling to the magnetic gate is tuned to be ``chiral". While there is no rectification of the magnon current in the film, we predict that chirality diverts a large percentage
(50$\%$ for perfect chirality) of it into the gate.  This transverse thermal spin pumping effect can be controlled by rotating the film magnetization and may help manage the heat flow in future magnonic circuits. 


\end{abstract}
\maketitle

\textit{Introduction}.---Thermal rectification implies the possibility to transport heat against a temperature gradient, a potentially important tool to modulate and manipulate the temperature in nanoscale structures \cite{heat_rectifier,Radiative_thermal_diode,review_rectification}. Rectification or diode effects in electronics and phononics \cite{heat_rectifier,heat_rectifier,Radiative_thermal_diode} are phenomena beyond linear response theory. Another phenomenon is the non-reciprocity or chirality in the excitation and propagation of coherent quasi-particles, \textit{i.e.} the coupling between their propagation direction and internal degrees of freedom, such as spin or orbital angular momentum \cite{chiral_optics,Yu_chirality,chiral_spintronics}.   The chiral nature of Damon-Eshbach spin waves on the surface of a ferromagnet \cite{DE}, for example, enables ``conveyor belt" transport of magnons in one direction \cite{heat_conveyor_1,heat_conveyor_2,heat_conveyor_3} irrespective of a temperature gradient. Non-reciprocity was also observed in non-local thermal transport experiments in thin yttrium iron garnet (YIG) films with Pt spin injection and detection as ascribed to an interfacial Dzyaloshinskii-Moriya interaction \cite{DMI_1,DMI_2} or an attached floating permalloy gate \cite{Liu_nonreciprocal}. Similar non-reciprocal magnon transport was reported in antiferromagnets as well \cite{nonreciprocal_Hanle_effect}. The blocking of one in favor of the other current direction can then be confused with rectification or diode effects.

\begin{figure}[ht]
\centering
\includegraphics[width=0.43\textwidth]{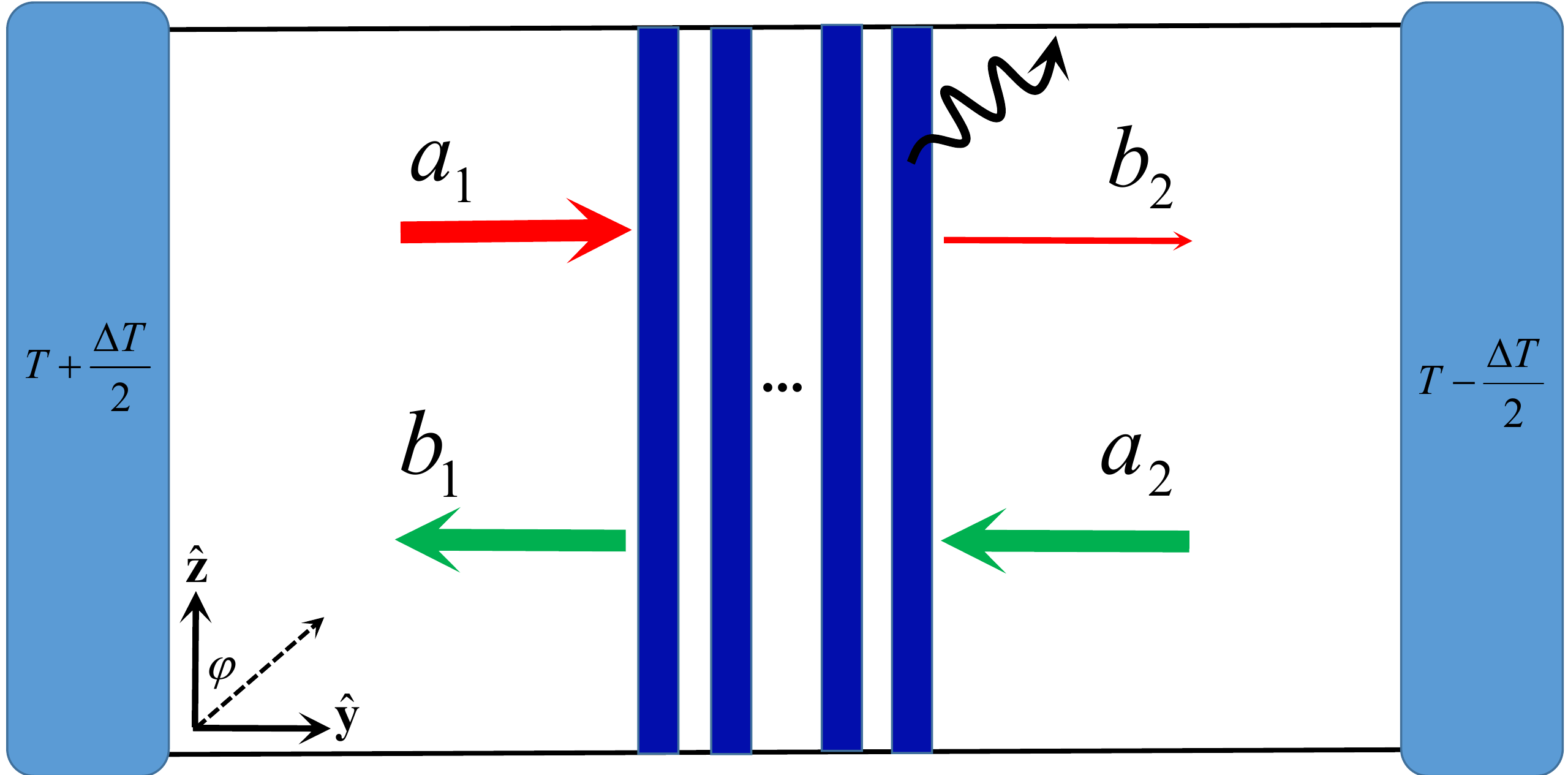}
\caption{The exchange of magnons between a magnetic film that carries a heat current and a proximity magnetic nanowires can be chiral, i.e., depending on the propagation direction of the magnons in the film.  $a_{1,2}$ are the input amplitudes of (spin) waves and $b_{1,2}$ are their output. The gate (the blue-striped region) can be tuned to modulate only the transmission amplitude $a_1$.}
\label{model}
\end{figure}

In this work, we address thermal transport in the linear regime for a generic magnon transistor~\cite{Bart_transistor,Huebl_transistor,Xiangyang_transistor}  composed of a YIG film that is partly covered by an array of magnetic nanowires, as sketched in  Fig.~\ref{model}. Baumgaertl and Grundler report that such a structure may serve as a non-volatile memory in which a magnon current in the YIG film switches the nanowire magnetization~\cite{magnetization_reversal}, which urgently calls for a microscopic model.  The dipolar interaction between quasiparticles in the film and wires is ``chiral", \textit{i.e.}, for a fixed magnetization it depends on the propagation direction~\cite{chiral_optics,Yu_chirality}, as widely predicted and observed for coherent waves in the form of, e.g., chiral pumping~\cite{Kruglyak_simulation,nanowire_Yu_1,nanowire_Yu_PRL,nanowire_Yu_exp,nanowire_Yu_exp_2}, chiral spin transport~\cite{ferroelectric}, non-reciprocal transmission of microwaves \cite{waveguide_Yu_PRB,waveguide_exp,waveguide_1963,waveguide_quantum_steering}, and the ``phonon diode"  \cite{phonon_diode_Xu,phonon_diode_Yu,phonon_diode_Wixforth,phonon_diode_Page}. These systems represent the   \textit{general} problem characterized by Hamiltonians with non-reciprocal coupling constant $|g_k|\ne |g_{-k}|$  between two different fields or particles \cite{Yu_chirality}. The thermodynamic consequences of chirality have yet often been overlooked, as reflected by the frequent but incorrect use of the word ``diode effect" or ``rectification" when the chirality is meant. The former is a relation  $J(\Delta T)\ne -J(-\Delta T)$  for a current \(J\) that when driven in linear response by a temperature bias $\Delta T$ would imply a (non-existent) autonomous Maxwell Demon or ``trapdoor". Thermal fluctuations in the temperature gradients and currents that average out to zero over time cause, for instance, the Johnson-Nyquist noise in electronic circuits~\cite{noise}. A relation $J(\Delta T)\ne -J(-\Delta T)$ would rectify the equilibrium thermal noise, in violation of the Second Law of Thermodynamics. 

In the linear response regime, the chiral coupling cannot cause rectification, so our emphasis lies on the elucidation of its observable physical consequences. To this end, we treat the (floating) magnetic nanowire gate as a third thermal bath that exchanges magnons with the film in a chiral manner, e.g. by causing damping of the waves in the film that propagate into one direction, but not those into the other. We formulate the problem in terms of self-consistent non-equilibrium Green functions and confirm that in the absence of a temperature difference between the source and drain contacts, all currents vanish on average. There is no rectification of the equilibrium currents \(J\) under a temperature bias \(\Delta T\), \textit{viz}.  \(J(\Delta T)=-J(-\Delta T)\).  Nevertheless, we predict that chirality causes a surprising and contra-intuitive effect: In the absence of chirality and a symmetric configuration, the floating gate is inert but switching on the chirality diverts a large percentage (one-half for perfect chirality)  of the magnon heat and spin currents into the gate. Since the chirality depends strongly on the magnetic configuration we can control the effect by a weak in-plane magnetic field, which might be useful for thermal management in magnonic circuits.

\textit{Model for a non-unitary scattering matrix}.---We illustrate the physics and our formalism at the hand of ballistic magnons in thin YIG films of thickness ${\cal O}(10~{\rm nm})$ that couple to metallic magnetic nanowires  \cite{Kruglyak_simulation,nanowire_Yu_1,nanowire_Yu_exp,nanowire_Yu_PRL,nanowire_array_PRL} or magnetic bi- and multilayers \cite{magnetic_film_1,magnetic_film_2,magnetic_film_3,magnetic_film_4} that when strongly damped act as the reservoirs in the scattering theory of transport  \cite{Landauer_1,Landauer_2}. More generally, they form a non-Hermitian dynamic ``scattering potential", in the adiabatic limit giving rise to spin pumping and current-induced magnetization dynamics. Here we address the effects of chiral coupling on dissipative and non-adiabatic transport. 

For computational convenience, we adopt a generic scattering region that consists of a regular array of floating metallic magnetic nanowires on top of the YIG film~\cite{nanowire_Yu_exp,nanowire_array_PRL} and along the $\hat{\bf z}$-direction with inter-wire distance $L_0$, counted by the index $l=\{1,\cdots, N\}$ as in Fig.~\ref{model}. A thin insulating layer such as ${\rm Al_2O_3}$ suppresses any interlayer exchange interaction, such that the magnetizations are only coupled by the magneto-dipolar interaction as observed in many recent experiments~\cite{nanowire_Yu_exp,Hanchen_two_wires,nanowire_Yu_exp_2,magnetization_reversal}. A strong external field along the wires first aligns the magnetizations.  For a YIG film, a \textit{weak}  ($\gtrsim$5 mT) in-plane applied magnetic field ${\bf H}_0$ freely rotates the substrate magnetization into any direction, while the shape anisotropy strongly pins the wire magnetization, as substantiated by model calculation in the Supplemental Material (SM).


Magnons in thin YIG films thermally populate the lower perpendicular standing spin waves (PSSWs), which is effectively a single one when the film is sufficiently thin and/or the temperature is sufficiently low \cite{Xiang_Yang}. The lowest PSSW  also couples most efficiently with adjacent magnetic nanowires. We defer the standard calculation of the magnon modes to the SM.
Hence, we focus on this two-dimensional limit with Hamiltonian for the film 
\begin{align}
\hat{H}_m/\hbar=\sum_{\kappa}\sum_{k_z}\omega_{\kappa}(k_z)\hat{m}^{\dagger}_{{\kappa}}(k_z)\hat{m}_{{\kappa}}(k_z),
\end{align}
where $\omega_{\kappa}(k_z)$ is the magnon dispersion relation and
$\hat{m}_{\kappa}(k_z)$ annihilates a magnon with wave vector ${\kappa}\hat{\bf y}+k_z\hat{\bf z}$. \(\hat{\bf y}\) is the transport direction.  The dipolar interaction between wires for inter-wire distances $L_0$ several times larger than their widths $w$ can safely be disregarded. The nanowires are thin and narrow, so their Hamiltonian reads
\begin{align}
\hat{H}_s/\hbar=\sum_l\sum_{k_z}\Omega({k_z})\hat{\beta}_{l}^{\dagger}(k_z)\hat{\beta}_{l}(k_z),
\end{align}
where $\Omega({k_z})$ is the frequency of a magnon $\hat{\beta}_l(k_z)$ in the $l$-th wire that propagates freely along the wire $\hat{\bf z}$-direction.
The coupling between the magnons in film and wires
\begin{align}
\hat{H}_c/\hbar=\sum_{l,k_z}\sum_{\kappa}g_{\kappa,l}({k_z})\hat{\beta}_{l}^{\dagger}(k_z)\hat{m}_{{\kappa}}(k_z)+{\rm H.c.},
    \label{interaction}
\end{align}
where the dipolar interaction $g_{\kappa,l}(k_z)$ is non-reciprocal, \textit{viz}. $|g_{\kappa,l}(k_z)|\ne |g_{-\kappa,l}(k_z)|$ \cite{nanowire_Yu_1}, referring for details  to the SM. 

For sufficiently large damping of the magnetization dynamics in the wires, any non-equilibrium state is dissipated quickly and cannot cause a back action on the film. The energy transfer from the film to the wires then breaks the unitarity of the scattering matrix of magnons in the film. Calling the conserved wave number along the long wires $k_z$, the scattering amplitude of film magnons from $\kappa\hat{\bf y}$ to $\kappa'\hat{\bf y}$ may be expressed by the $T$-matrix \cite{scattering_PRB,scattering_PRE,Mahan}
 \begin{align}
 T_{{\kappa}'{\kappa}}(k_z)&=\delta_{{\kappa}'{\kappa}}+\frac{{\cal M}_{{\kappa}'}^{\dagger}(k_z)
 {\cal G}^r_{k_z}(\omega_{\kappa,k_z})
 {\cal M}_{\kappa}(k_z)}{\omega_{\kappa}(k_z)-\omega_{{\kappa}'}(k_z)+i0_{+}},
 \label{T_matrix}
 \end{align}
 where ${\cal M}_{\kappa}(k_z)=\begin{pmatrix}g_{\kappa,1}(k_z),\cdots,g_{\kappa,N}(k_z)
 \end{pmatrix}^T$  and the matrix ${\cal G}^r_{k_z}(\omega)=\left(\omega-{\cal H}_{k_z}(\omega)\right)^{-1}$ is the (retarded) Green function of the non-Hermitian wire Hamiltonian \cite{Mahan,Fetter}
 \begin{equation}
 {\cal H}_{k_z}(\omega)|_{ll'}=\Omega({k_z})(1-i\alpha_G)\delta_{ll'}+\Sigma_{k_z}(\omega)|_{ll'}.
 \label{non_Hermitian_Hamiltonian}
 \end{equation}
Its damping Gilbert constant $\alpha_G$ can easily be engineered to be large ${\cal O}(0.01-0.1)$. $\Sigma$ is the self-energy matrix of the wire magnons
 \begin{align}
 \Sigma_{k_z}(\omega)|_{ll'}=\sum_{\kappa'} \frac{g_{\kappa',l}(k_z)g^*_{\kappa',l'}(k_z)}{\omega-\omega_{\kappa'}(k_z)+i0_+}.
 \end{align}
We disregard the damping of the film magnons, which is allowed for a material such as YIG with Gilbert constant much smaller than the wire $\alpha_G$.

The transmission and reflection amplitudes that make up the magnon scattering matrix $S$ for given $\kappa>0$ reads according to Eq.~(\ref{T_matrix})   \cite{scattering_PRB,scattering_PRE,Mahan}
\begin{align}
 \nonumber
 &S_{21}(\kappa,k_z)=1-\frac{iL_y}{|v_{\kappa}(k_z)|}
 {\cal M}_{\kappa}^{\dagger}(k_z)
 {\cal G}^r_{k_z}(\omega_{\kappa,k_z})
 {\cal M}_{\kappa}(k_z),\\
 &S_{11}(\kappa,k_z)=-\frac{iL_y}{|v_{\kappa}(k_z)|}
 {\cal M}_{-\kappa}^{\dagger}
 {\cal G}^r_{k_z}(\omega_{\kappa,k_z}){\cal M}_{\kappa}(k_z),
 \label{magnon_transmission}
 \end{align}
where $v_{\kappa}(k_z)=\partial \omega_{\kappa}(k_z)/\partial \kappa$ is the group velocity along the transport $\hat{\bf y}$-direction and $L_y$ is the sample length along $\hat{\bf y}$. We disregard a constant propagation phase factor in  $S_{21}(\kappa,k_z)$. $S_{12}(\kappa,k_z)$ and $S_{22}(\kappa,k_z)$ are obtained by replacing $|\kappa|\rightarrow -|\kappa|$ in Eq.~(\ref{magnon_transmission}). In the chiral limit, the reflections vanish with either $g_{\kappa,l}(k_z)=0$ or $g_{-\kappa,l}(k_z)=0$.
This scattering matrix is unitary $|S_{21}(\kappa,k_z)|^2+|S_{11}(\kappa,k_z)|^2=1$ only when the magnon number is conserved, \textit{i.e.}, when $\alpha_G\rightarrow 0$. In the limit of a single narrow wire, we can treat the problem analytically with
\begin{align}
\nonumber
    S_{21}(\kappa,k_z)&=1-\frac{i\Gamma_{R}(\kappa,k_z)}{\omega_{\kappa,k_z}-\Omega_{k_z}+i \left[\alpha_G\Omega_{k_z}+(\Gamma_{L}+\Gamma_{R})/{2}\right]},\\
    S_{12}(\kappa,k_z)&=1-\frac{i\Gamma_{L}(\kappa,k_z)}{\omega_{\kappa,k_z}-\Omega_{k_z}+i \left[\alpha_G\Omega_{k_z}+(\Gamma_{L}+\Gamma_{R})/{2}\right]},
    \nonumber
\end{align}
where $\Gamma_{L}(\kappa,k_z)={L_y} g^2_{-|\kappa|}(k_z)/{v_{\kappa}(k_z)}$ and $\Gamma_{R}(\kappa,k_z)={L_y} g^2_{|\kappa|}(k_z)/{v_{\kappa}(k_z)}$. In the chiral limit $\Gamma_L=0$,  the transmission $S_{12}$  for a single wire is unity, while the reflection $S_{22}$ vanishes, as illustrated by Fig.~\ref{model}.

In the following, we present numerical results for an array of $N=10$ nanowires with dimensions used in experiments  Refs.~\cite{nanowire_Yu_exp,nanowire_array_PRL}.  We calculate the retarded Green function ${\cal G}^r_{k_z}$ and scattering matrix Eq.~(\ref{magnon_transmission})  by inverting the non-Hermitian Hamiltonian (\ref{non_Hermitian_Hamiltonian}), referring to the SM for details.   We choose Co nanowires with width  $w=100$~nm, thickness $d=30$~nm, magnetization $\mu_0\tilde{ M}_s=1.1$~T, and exchange stiffness $\tilde\alpha_{\rm ex}=3.1\times10^{-17}$~${\rm m}^2$.  The YIG film is $s=10$~nm thick, with magnetization $\mu_0{M}_s=0.2$~T, and exchange stiffness $\alpha_{\rm ex}=3.0\times10^{-16}$~${\rm m}^2$ \cite{nanowire_Yu_exp,nanowire_array_PRL}. Moreover, $\mu_0H_0=0.05$~T, $L_0=300$~nm between the neighboring wires, the Co damping coefficient $\alpha_G=0.05$,  $\mu_0=4\pi\times10^{-7}$ $\rm H/m$, and $\gamma=1.82\times10^{11}$ $\rm s^{-1} T^{-1}$. 

\begin{figure}[htp]
\centering
\hspace{-0.2cm}
\includegraphics[width=0.23\textwidth,trim=1.2cm 0.8cm 2.4cm 0cm]{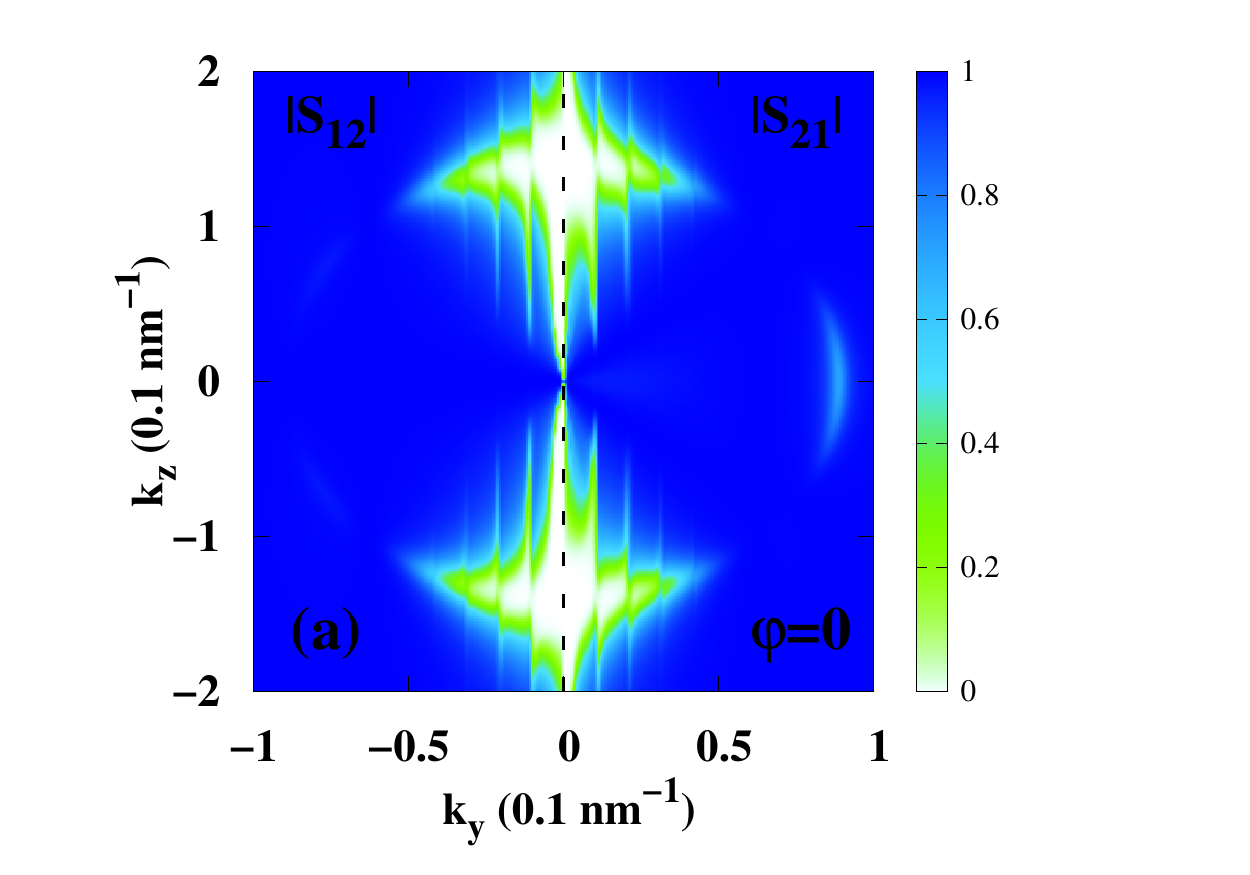}
\includegraphics[width=0.23\textwidth,trim=1.2cm 0.8cm 2.4cm 0cm]{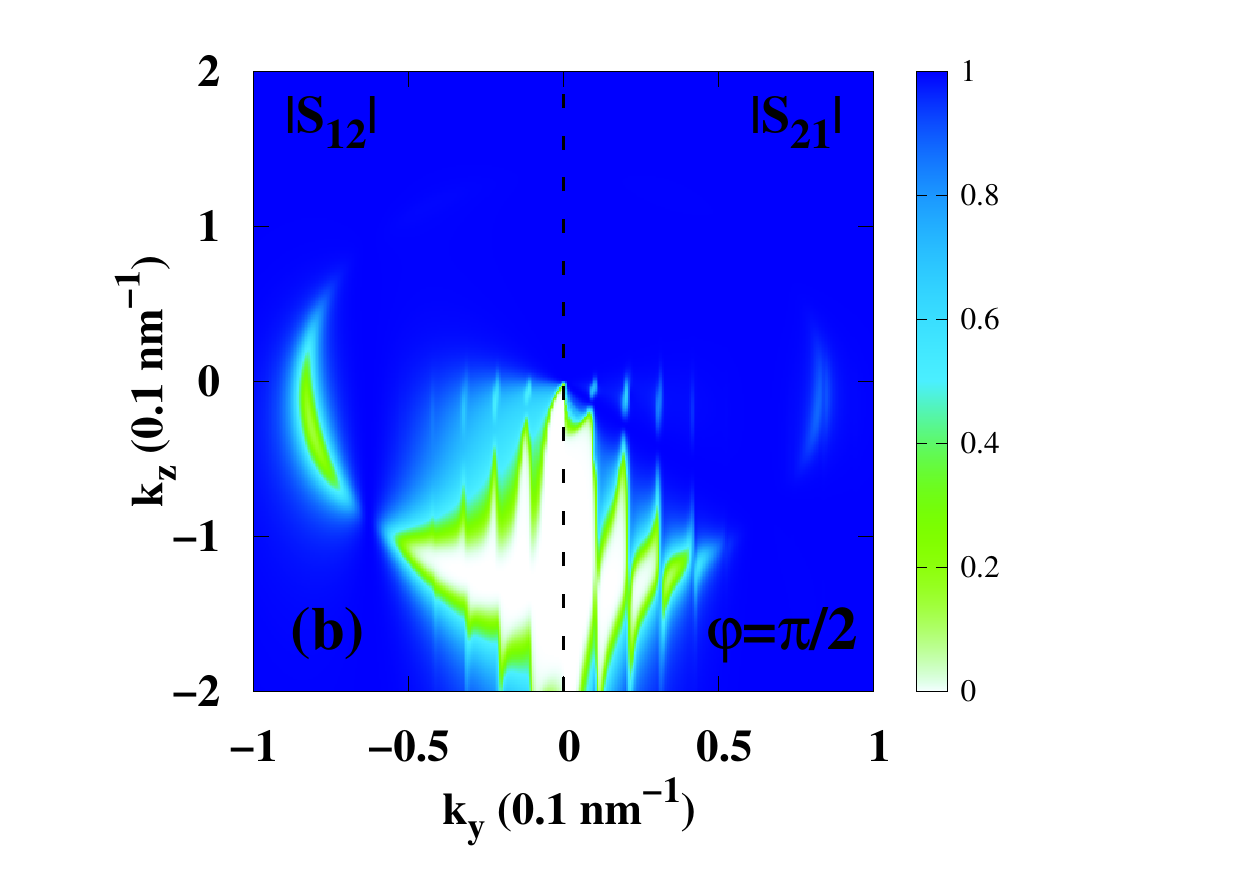}\\
\includegraphics[width=0.23\textwidth,trim=1.2cm 0.8cm 2.4cm 0cm]{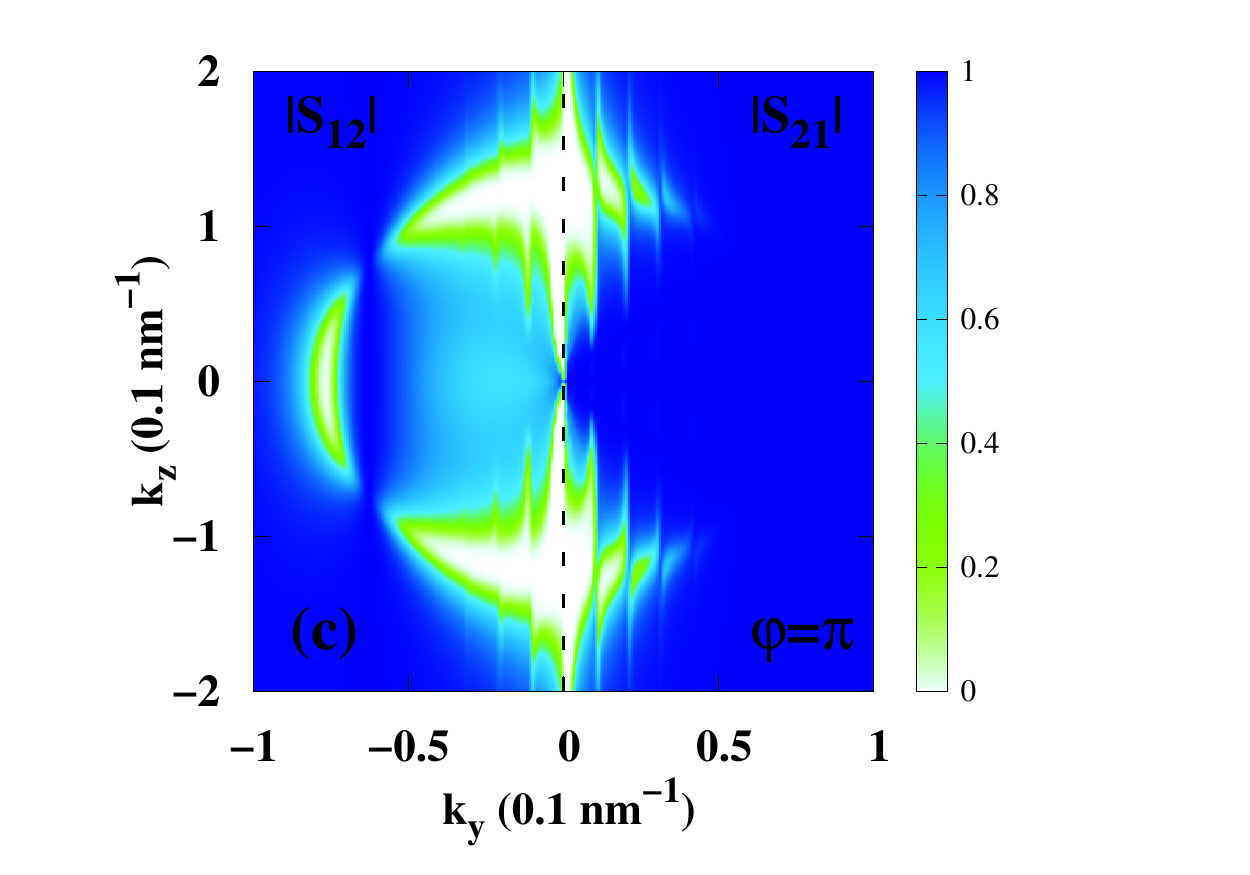}
\includegraphics[width=0.23\textwidth,trim=1.3cm 0.8cm 2.3cm 0cm]{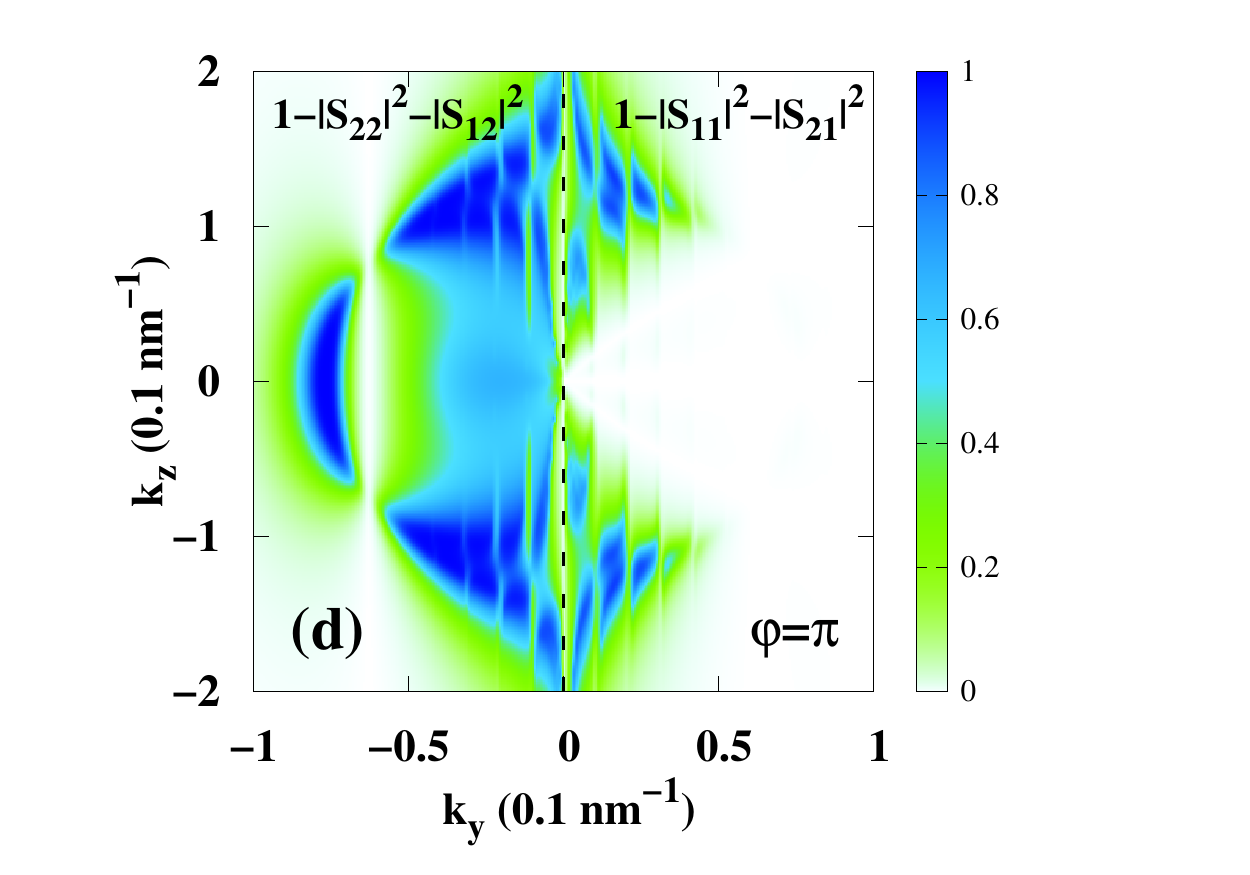}
\caption{Transmission coefficients $\{|S_{12}({\bf k})|, |S_{21}({\bf k})|\}$ and non-unitarity of the scattering matrix of spin waves in a YIG film as a function of wave vector ${\bf k}$  launched to the gate formed by a Co nanowire array for different magnetic configurations. (a)-(c) shows nearly total reflections when the film magnetization is parallel, normal, and antiparallel to that of the wires.  (d) shows the deviation of ${|S_{11}({\bf k})|^2+|S_{21}({\bf k})|^2}$ from unity, which is caused by the Gilbert damping in the wires. The associated non-unitarity is chiral with respect to $k_y$.}
\label{reflections}
\end{figure}

Figures~\ref{reflections}(a)-(c) show the moduli of the transmission amplitudes $|S_{12}({\bf k})|$ ($|S_{21}({\bf k})|$) of spin waves in the YIG film with wave vectors ${\bf k}$ that are launched from the left (right) to interact with the Co nanowires that form an angle  $\varphi$ of the film magnetization with the wire $\hat{\bf z}$-direction.  Figure~\ref{reflections}(d) express the non-unitarity of the scattering matrix when the magnetizations are antiparallel.  Figures~\ref{reflections}(a)-(c) reveal strong inter-magnet interactions  over a wide range of wavelengths of up to $\sim 50$~nm and deviations from the ideal chiral limit. The  non-reciprocity with respect to $k_y$ is pronounced in the antiparallel $\varphi=\pi$ configuration, \textit{i.e.},  when the dipolar fields of the spin waves in the film match those of the wire dynamics, as sketched in the SM. The damping in the Co wires breaks the unitarity of the scattering matrix, as emphasized by Fig.~\ref{reflections}(d). The non-unitarity is also chiral, almost perfectly so for $k_z=0$.

The non-reciprocity of the scattering matrix can be measured by spatially resolved Brillouin light scattering~\cite{BLS} and NV center magnetometry~\cite{NV_center,Toeno} of the spin wave stray fields. It also uniquely affects thermal magnon transport as demonstrated in the following.

\textit{Chirality-enabled thermal spin pumping.}---Here we address the nontrivial role of chirality in the thermal spin pumping into the magnetic nanowire. A temperature difference \(T_{S/D}=T_0 \pm \Delta T/2\)  between the source ``S" and drain ``D" reservoirs as shown in Fig.~\ref{model} drives heat and magnon currents in the YIG film.  We formulate the magnon transport in our dissipative system in terms of self-consistent non-equilibrium Green functions. The loss rate of  magnons in the film equals the difference in the rates of change of the ensemble averages $\langle\cdots\rangle$ in the presence and absence of the interaction Eq.~(\ref{interaction}). 

The density operator $\hat{\rho}_{\kappa}(k_z)=\hat{m}^{\dagger}_{\kappa}(k_z)\hat{m}_{\kappa}(k_z)$ obeys the Heisenberg equation of motion
\begin{align}
\nonumber
    \left\langle \frac{d\hat{\rho}_{\kappa}(k_z)}{dt}\right\rangle&=\frac{i}{\hbar}
    \left\langle \left[\hat{H}_m+\hat{H}_c,\hat{\rho}_{\kappa}(k_z)\right]\right\rangle\\
    &=2{\rm Re}\left(\sum_lg_{\kappa,l}^*({k}_z)G_{\kappa,l}^<(k_z;t,t)\right).
    \label{rate_of_change}
\end{align}
The transfer of magnons into the wires is therefore proportional to the coupling strength and the ``lesser" Green function $G_{{\kappa,}l}^<(k_z;t,t')\equiv -i\left\langle \hat{\beta}^{\dagger}_{l}(k_z,t')\hat{m}_{{\kappa}}(k_z,t)\right\rangle$. 
In frequency space \cite{Haug,Abrikosov}, 
$G_{\kappa,l}^<(k_z,\omega)=\sum_{l'}g_{\kappa,l'}(k_z)[{\cal G}^r_{k_z,ll'}(\omega){G}^{(0)}_{\kappa,<}(k_z,\omega)+{\cal G}_{k_z,ll'}^<(\omega){G}_{\kappa,a}^{(0)}(k_z,\omega)]$,
where for magnons in the unperturbed film  ${G}^{(0)}_{\kappa,<}(k_z,\omega)=2\pi i f(\omega,T)\delta\left(\omega-\omega_{\kappa,k_z}\right)$ and the ``advanced'' Green function
${G}^{(0)}_{{\kappa},a}(k_z,\omega)={1}/({\omega-\omega_{\kappa,k_z}-i0_+})$ with Planck distribution $f(\omega,T)=1/\{\exp[\hbar\omega/(k_BT)]-1\}$ at temperature $T$. On the other hand, when the Gilbert damping and magnon-magnon interaction in the wires is strong,  their magnon distribution is at equilibrium with the ambient temperature $T_0$, allowing for exploiting the equilibrium Green function ${\cal G}_{k_z}^<(\omega)=-f(\omega,T_0)[{\cal G}^r_{k_z}(\omega)-{\cal G}^{r\dagger}_{k_z}(\omega)]$ \cite{Haug,Abrikosov}. 
Substituting this into Eq.~(\ref{rate_of_change}), we arrive at
\begin{align}
\nonumber
    &\left\langle {d\hat{\rho}_{\kappa}(k_z)}/{dt}\right\rangle\\
    &={\rm Re}\big[i{\cal M}^{\dagger}_{\kappa}(k_z){\cal G}_{k_z}^r(\omega_{\kappa,k_z}){\cal M}_{\kappa}(k_z)\big]\nonumber\\
    &\times\left(f(\omega_{\kappa,k_z},T)-f(\omega_{\kappa,k_z},T_0)\right)\nonumber\\
    &+{\rm Re}\big[i{\cal M}^{\dagger}_{\kappa}(k_z){\cal G}_{k_z}^r(\omega_{\kappa,k_z}){\cal M}_{\kappa}(k_z)\big]f(\omega_{\kappa,k_z},T)\nonumber\\
    &+{\rm Re}\big[i{\cal M}^{\dagger}_{\kappa}(k_z){\cal G}_{k_z}^{r\dagger}(\omega_{\kappa,k_z}){\cal M}_{\kappa}(k_z)\big]f(\omega_{\kappa,k_z},T_0).
    \label{rate_of_change_2}
\end{align}
The terms associated with the distribution $f(T)$ account for the spin injection from films to wires, which is balanced by the inverse processes proportional to $f(T_0)$. When $T=T_0$,
$\left\langle {d\hat{\rho}_{\kappa}(k_z)}/{dt}\right\rangle={\rm Re}[i{\cal M}^{\dagger}_{\kappa}(k_z)({\cal G}_{k_z}^{r}(\omega_{\kappa,k_z})+{\cal G}_{k_z}^{r\dagger}(\omega_{\kappa,k_z}))
{\cal M}_{\kappa}(k_z)]f(\omega_{\kappa,k_z},T_0)=0$,
so the scattering vanishes as required by the principle of detailed balance, as do equilibrium magnon currents, in spite of the chirality of the interaction.

Next, we relate the magnon transmission Eq.~(\ref{magnon_transmission}) to the spin-injection rates of magnons with $\kappa\hat{\bf y}+k_z\hat{\bf z}$.  The magnons in the film are weakly perturbed such that  $\delta f_{\kappa}(k_z)=f(\omega_{\kappa,k_z},T)-f(\omega_{\kappa,k_z},T_0)$ is small. Here we distinguish the wave numbers $\kappa_L$ and $\kappa_R$ in the left ``L" and right ``R" reservoirs. Under the temperature bias
\begin{align}
    \delta f_{\kappa}(k_z)=\frac{\partial f(\omega_{\kappa,k_z},T)}{\partial T_0}\left\{
    \begin{array}{c}
     \Delta T/2,~~~~\kappa\in L\\
     -\Delta T/2,~~~\kappa\in R
    \end{array}\right.,
\end{align}
magnons with $\kappa_L>0$ are absorbed by the nanowires and after thermalization re-emitted to travel into the right reservoir at the same rate.
The spin-injection rate from different reservoirs relates to the magnon transmission \eqref{magnon_transmission} as
 $\left\langle d\hat{\rho}_{\kappa,k_z}/dt\right\rangle={\delta f_{\kappa}(k_z)}{\Gamma_{\kappa}(k_z)}$,
\textit{viz.}, the wires dissipate magnon energy at the rates 
\[
    \Gamma_{\kappa}(k_z)=\frac{|v_{\kappa}(k_z)|}{L_y}\left\{
    \begin{array}{c}
     1-{\rm Re}S_{21}(\kappa,k_z),~~~\kappa_{L,R}>0\\
     1-{\rm Re}S_{12}(\kappa,k_z),~~~\kappa_{L,R}<0
    \end{array}\right.,
\]
which vanish when the transmission is high but become significant $\Gamma_{\kappa}(k_z)\rightarrow |v_{\kappa}(k_z)|/L_y$ otherwise. In the chiral limits ${\rm Re}S_{21}\rightarrow -1$ or ${\rm Re}S_{12}\rightarrow -1$, the injection rate $\Gamma_{\kappa}(k_z)$ is enhanced by a factor 2 because the source injects magnons into the wires while the drain contributes magnon ``holes".

Exchange magnons carry an angular momentum $\hbar$ and energy $\hbar\omega_{\kappa}(k_z).$ The spin [heat] current density flowing from the film into the gate then read

\begin{align}
J_{s[E]} &= \frac{\hbar}{L_z} \sum_{\kappa}\sum_{k_z}{\delta f_{\kappa}(k_z)}\big[\omega_{\kappa}(k_z)\big]{\Gamma_{\kappa}(k_z)} \nonumber \\
    &=\frac{\hbar}{L_yL_z}\frac{\Delta T}{2}\sum_{\kappa>0,k_z}\frac{\partial f(\omega_{\kappa,k_z})}{\partial T_0}\big[\omega_{\kappa}(k_z)\big] \nonumber \\
    &\times|v_{\kappa,k_z}|\big({\rm Re}S_{12}(-\kappa,k_z)-{\rm Re}S_{21}(\kappa,k_z)\big).
    \label{currents}
\end{align}
These expressions do not depend on a sufficiently large width $L_z$.
The right- and left-moving spin waves are distributed according to the temperature of their source. The hot magnons from the source are annihilated when injected into the nanowires.  Without chirality ${\rm Re}\,S_{12}(-|\kappa|,k_z)={\rm Re}\,S_{21}(|\kappa|,k_z)$, the in- and out-going magnon currents cancel to $J_{s[E]}=0$. However, a finite chirality Eq.~(\ref{currents})  leads to heating \textit{vs}. cooling of the wire, depending on the coupling parameter that can be tuned by the in-plane magnetic field and the sign of the applied temperature bias.

The transistor can divert maximally $1/2$ of the spin current into the nanowires when the chiral damping is complete such that all magnons from one direction are fully dissipated, while those flowing in the other direction are not affected at all,  i.e., $[{\rm Re}S_{12}(\kappa,k_z)-{\rm Re}S_{21}(-\kappa,k_z)]=1$ for all $k_z$, leading to the spin current injected into wires
$J_s=\hbar({\Delta T}/2)\int_0^{\infty}d\kappa\int dk_z|v_{\kappa,k_z}|{\partial_{T_0}} f(\omega_{\kappa,k_z})$. On the other hand, the flow in the substrate contains the contribution from the right reservoir that is the same as $J_s$, so the current flowing in the substrate  $J_L= 2J_s$.

\begin{figure}[htp]
\centering
\hspace{-0.23cm}
\includegraphics[width=0.242\textwidth,trim=0.6cm 1.1cm 0cm 0.5cm]{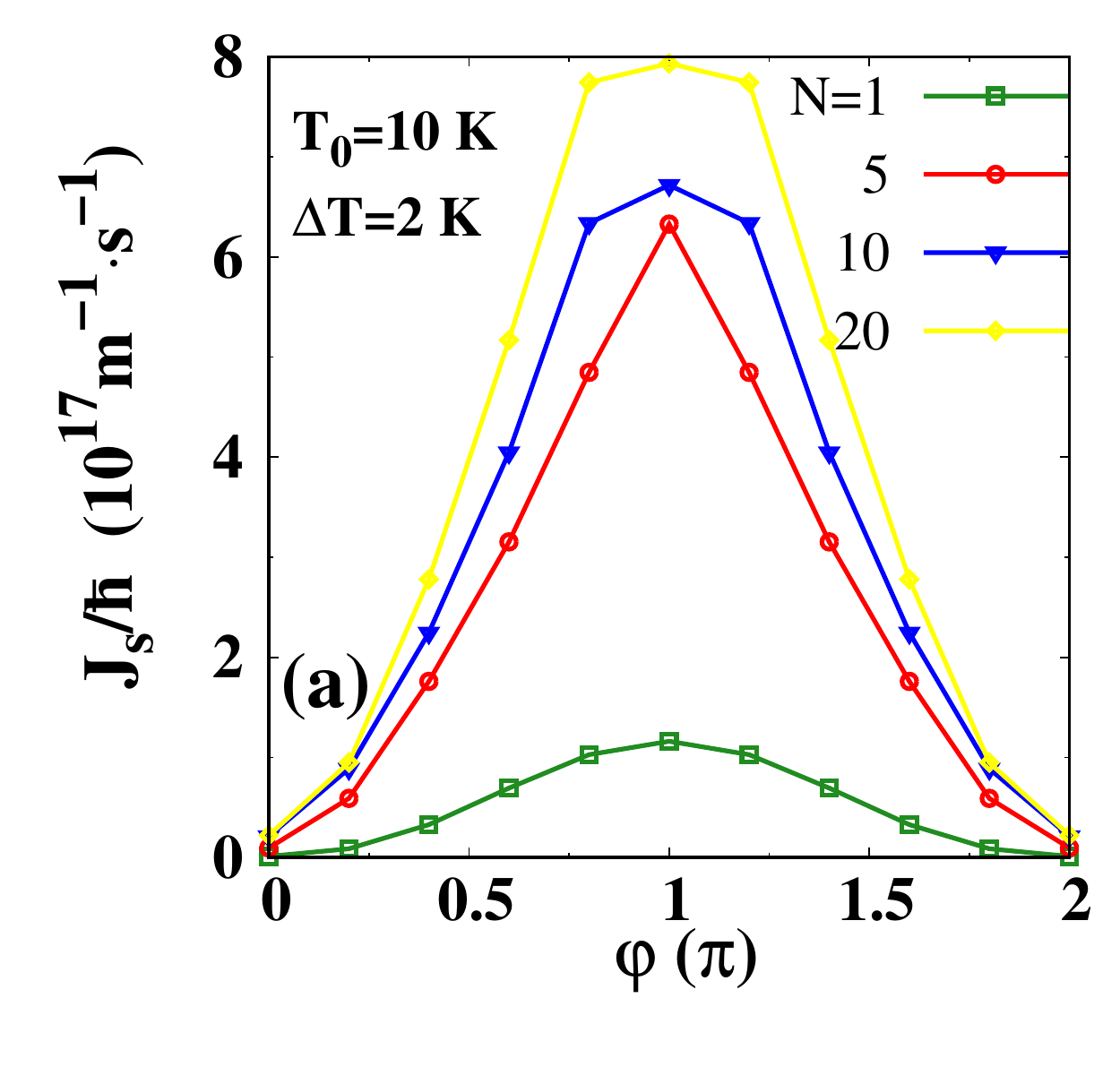}
\includegraphics[width=0.242\textwidth,trim=0.6cm 1.1cm 0cm 0.5cm]{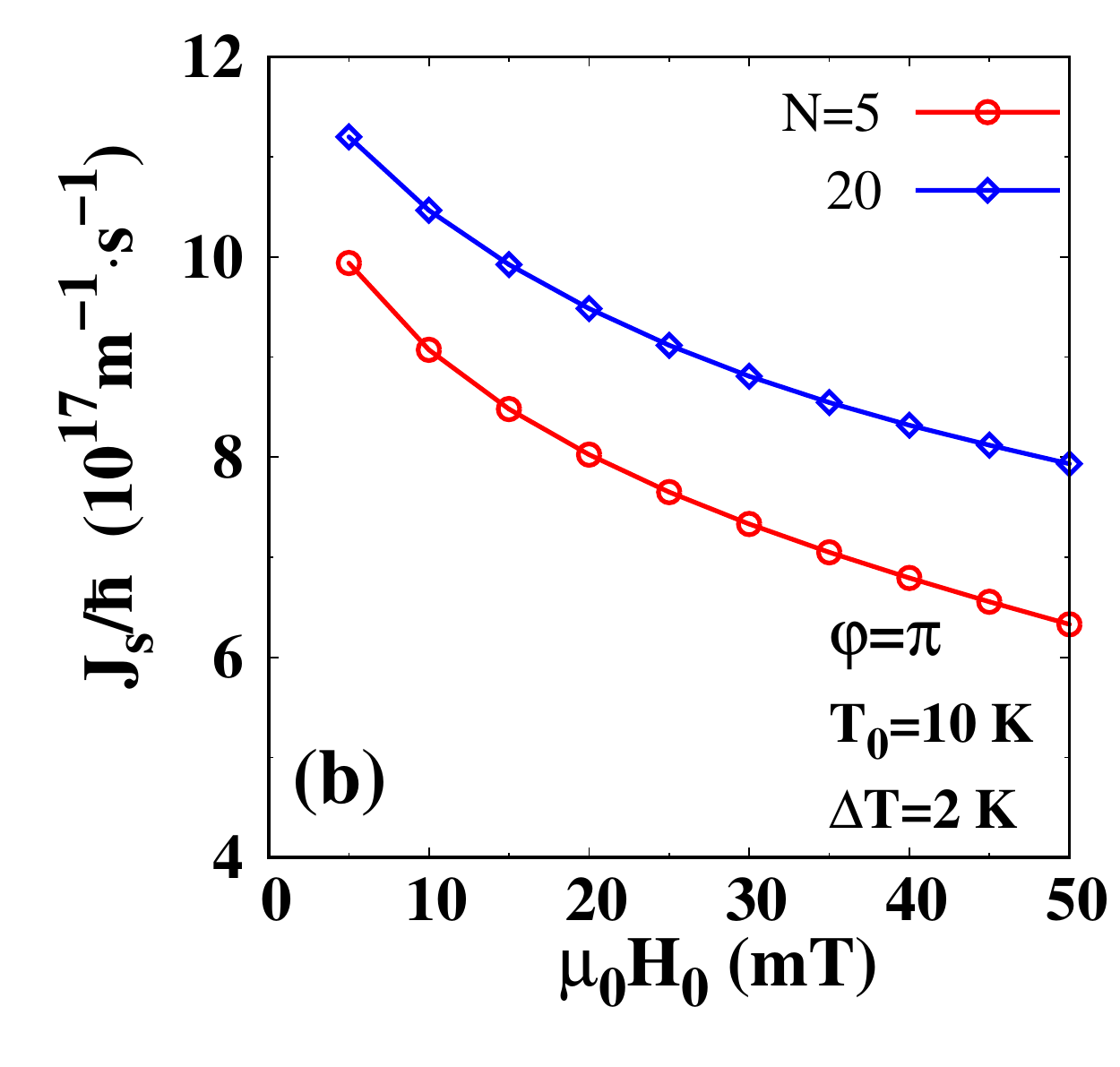}\\
\includegraphics[width=0.23\textwidth,trim=1.3cm 0cm 0cm 0cm]{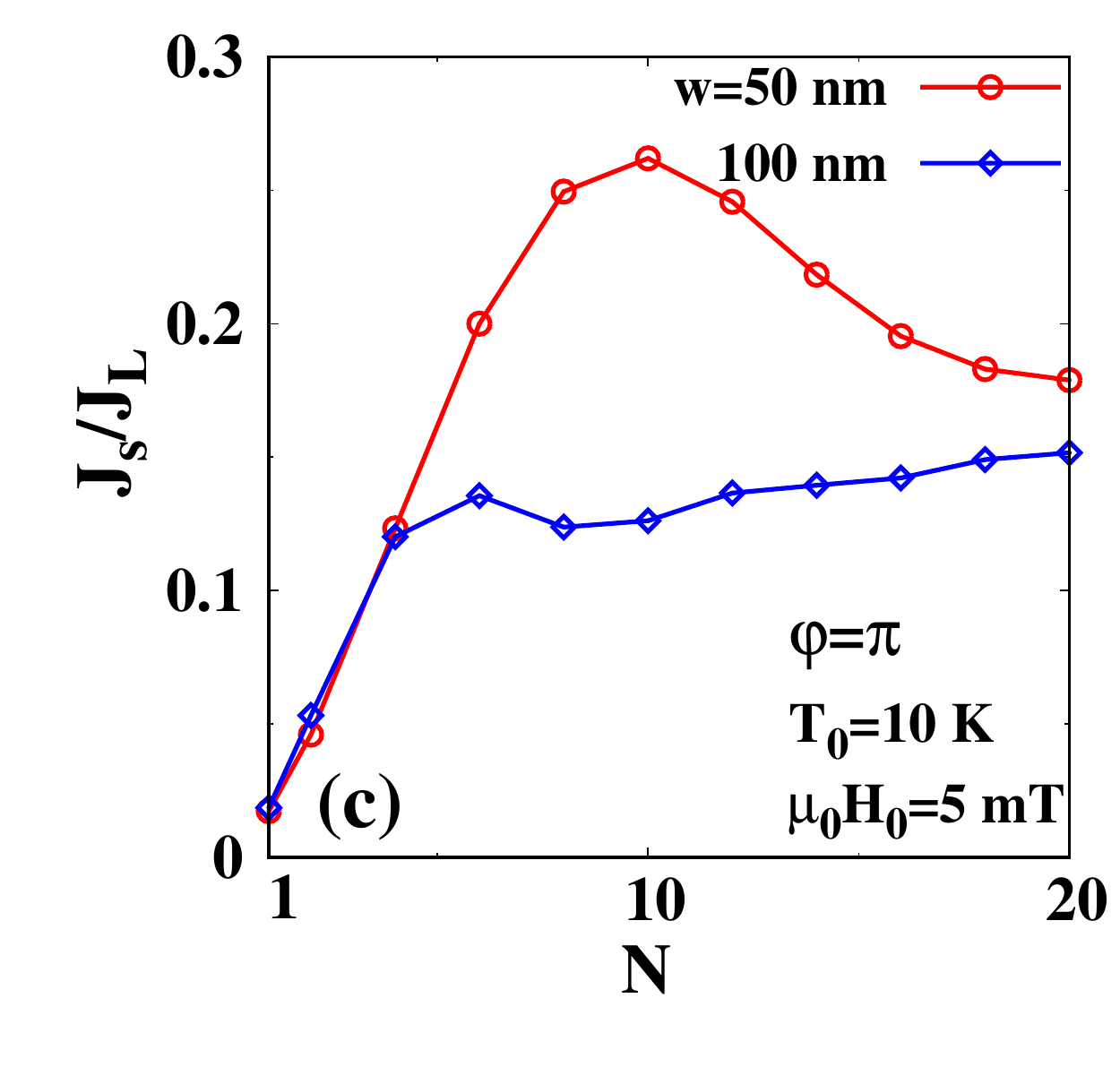}
\includegraphics[width=0.236\textwidth,trim=0.7cm 0cm 0.3cm 0cm]{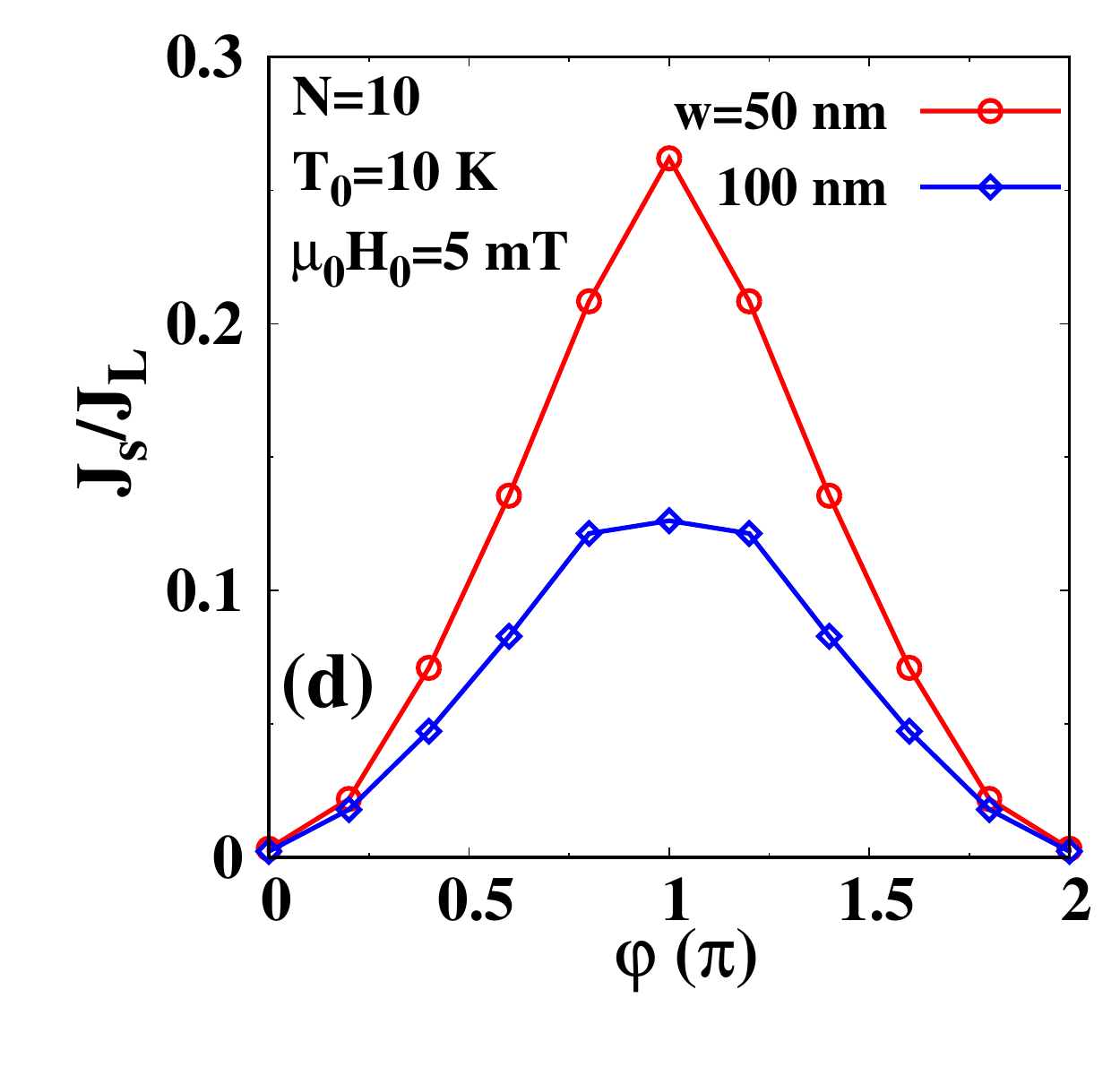}
\caption{Spin current $J_s$ channeled off into the gate from the source to drain [(a) and (b)] and its modulation efficiency $J_s/J_L$ [(c) and (d)] as a function of magnetization direction $\varphi$, the external magnetic field, and the number of magnetic wires $N$.}
\label{spin_current}    
\end{figure}

Figure~\ref{spin_current} plots the spin current dissipated by the nanowires as a function of magnetization direction $\varphi$ of the film as well as magnetic wire number $N$ when $T_0=10$~K and $\Delta T=2$~K. The magnetization direction modulates the chirality and thereby the thermal spin pumping into the nanowires by two orders when rotated from a parallel $(\varphi=0)$ to an antiparallel $(\varphi=\pi)$ configuration (Fig.~\ref{reflections}).
Increasing the number of wires from 1 to 20 at a constant inter-wire distance enhances the effect by an order of magnitude. 
In the YIG/Co system, \(J_L/\hbar\) is typically $1.5\times10^{19}$~${\rm m}^{-1}{\rm s}^{-1}$ for $\Delta T=2$~K and $T_0=10$~K. We compute a modulation efficiency \(J_s/J_L\)  of $\sim 13\%$ ($\sim 27\%$) for $N=10$ magnetic wires of width $w=100$~nm (50~nm) and $\mu_0H_0=5$~mT with magnetizations opposite to that of the substrate, which is not far from the ideal maximal efficiency [Fig.~\ref{spin_current}(c) and (d)].

The spin injection from the substrate into nanowires enabled by the chirality provides an anti-damping torque on the nanowire magnetization, which is the largest in the antiparallel configuration. The effect might allow switching by temperature gradient and explain the observed magnetization reversal~\cite{magnetization_reversal}, but more work is needed to understand the magnitude of the effect.

\textit{Diverting heat and spin flows by chirality}.---The above result corresponds to a magnon transistor action \cite{Bart_transistor,Huebl_transistor,Xiangyang_transistor}. Even though the gate is not biased, it controls the thermal magnon current from the high-temperature contact into the gate by the angle  $\varphi$. This can be understood in terms of the Landauer-B\"uttiker formula \cite{Landauer_1,Landauer_2} for a three-terminal device, in which the gate is modeled as a thermodynamic reservoir (refer to the SM for a derivation). In this formalism, all dissipation occurs only in the reservoir, while the Hamiltonian of the scattering region is  Hermitian. The spin current leaving the source then reads
\begin{align}
\nonumber
J_L&=\frac{\hbar}{L_yL_z}\sum_{\kappa>0,k_z}
v_{\kappa,k_z}\left(1-\sum_{\kappa'>0}\left|T_{-\kappa\kappa'}(k_z)\right|^2\right)\delta f_L(\omega_{\kappa,k_z})\\
    &-
    \frac{\hbar}{L_yL_z}\sum_{\kappa>0,k_z}v_{\kappa,k_z}\left(\sum_{\kappa'<0}\left|T_{-\kappa\kappa'}(k_z)\right|^2\right)\delta f_R(\omega_{\kappa,k_z}),\nonumber
\end{align}
which vanishes at equilibrium $f_L=f_R=f(T_0)$,  even when chirality fully suppresses the magnon propagation in one direction. 
Moreover, under a temperature bias
$\delta f_{L/R}(\omega_{\kappa,k_z})=({\partial f(\omega_{\kappa,k_z},T)}/{\partial T_0})(+/-){\Delta T}/{2}$, $J_L(\Delta T)=-J_L(-\Delta T)$, so there is no rectification of the thermal current, either. In the chiral limit, these results confirm that optimally  $J_L=\hbar{\Delta T}\int_0^{\infty}d\kappa\int dk_z|v_{\kappa,k_z}|{\partial_{T_0}} f(\omega_{\kappa,k_z})=2J_s$. This result cannot explain the non-reciprocity observed in non-local thermal transport experiments in thin YIG films with Pt spin injection and detection under a floating permalloy gate~\cite{Liu_nonreciprocal}. 

\textit{Conclusion}.---We predict a non-local thermal spin pumping or spin Seebeck effect that only appears in the presence of chiral damping in gated magnetic nanostructures. Via tuning the direction of the magnetization in the film,  a large percentage
(50$\%$ for perfect chirality) of the heat and spin currents in the film can be diverted into the magnetic nanowires, a functionality that may help manage heat in magnonic circuits. The spin injection may help explain the observed magnetization reversal by magnon currents. Our results may be relevant for other chirally coupled non-Hermitian systems, such as chiral optics \cite{chiral_optics}, phononics, and plasmonics \cite{Yu_chirality}.

\begin{acknowledgments}
This work is financially supported by the National Key Research and Development Program of China under Grant No.~2023YFA1406600, the National Natural Science Foundation of China under Grant No.~12374109, the startup grant of Huazhong University of Science and Technology, as well as JSPS KAKENHI Grants No.~19H00645 and 22H04965. We thank Prof.~Jing-Tao L\"u for
valuable discussions.
\end{acknowledgments}

\end{document}